\documentclass[aps,footinbib,showpacs]{revtex4-1}

\usepackage{amssymb,latexsym,mathrsfs}
\usepackage{amsfonts}
\usepackage{graphicx}
\usepackage{color}    
\usepackage{textcomp}

\newcommand{\be}{\begin{equation}}
\newcommand{\ee}{\end{equation}}
\newcommand{\bearr}{\begin{array}}
\newcommand{\enarr}{\end{array}}

\newcommand{\ra}{\rangle}
\newcommand{\la}{\langle}

\def\bea{\begin{eqnarray}}
\def\eea{\end{eqnarray}}
\def\ba{\begin{array}}
\def\ea{\end{array}}

\begin{document}

\title{Assisted exchange models in one dimension}
\author{Amit  Kumar Chatterjee}\email{amit.chatterjee@saha.ac.in}  \author{P. K. Mohanty} 
\affiliation {CMP Division, Saha Institute of Nuclear Physics, HBNI,1/AF Bidhan Nagar, Kolkata, 700064 India.}


\begin{abstract}
We introduce an {\it assisted} exchange model (AEM)on a one dimensional periodic lattice with $K+1$ different species of hard core particles, where the exchange rate depends on the pair of  particles  which  undergo exchange and their  immediate left neighbor. We show that this stochastic process has  a pair factorized steady state for a broad class of  exchange dynamics.   We calculate exactly the particle current and  spatial correlations $(K+1)$-species AEM using a  transfer matrix formalism. Interestingly the current in AEM exhibits  density dependent current reversal and negative differential mobility- both of which have been discussed elaborately by using a two  species exchange  model which  resembles  the  partially asymmetric conserved lattice gas model  in one dimension. Moreover,  multi-species version of AEM exhibits  additional features like multiple points of current reversal,  and  unusual  response  of particle current.
\end{abstract}
\maketitle

\section{Introduction}
\label{intro}
Non-equilibrium phenomena \cite{book1} are very common in nature and occur whenever there is a net unbalanced flux of some physical observable like mass, charge etc. along some particular direction- this net flow of physical quantities, commonly termed 
as {\it current} is one of the most intriguing feature that characterizes non-equilibrium systems(NES) in contrast to their 
equilibrium counterparts which are identified as zero-current states. While the stationary states of equilibrium systems 
are quantified by the very well known Gibbs measure or Boltzmann distribution, the steady state measure of non-equilibrium 
systems has no unique identity and has always been a question of interest in the quest of NES. In this context, finding 
current or net flux in these non-equilibrium steady states and studying several features of the current in order to 
understand and characterize non-equilibrium systems has become one of the primary interests. For example, the exact large deviation 
function of the time averaged current resembling to the pressure of an ideal Bose or Fermi gas is obtained in \cite{cur_press}, anomalous 
behavior of cumulants of current near phase transition is shown in \cite{cfluc_phse_tran}, universal behavior of current independent of the dimension 
in symmetric exclusion process connected to reservoirs and its possible generalization to other diffusive systems is discussed in 
\cite{cur_universal}. Also, interesting features like reversal of direction of current with respect to density \cite{crev_azrp} and 
negative differential mobility \cite{ndm_1,ndm_2,ndm_3,ndm_4}
and absolute negative mobility \cite{anm_1,anm_2,anm_3} in interacting particle systems has caused much attention now a days.

Driven diffusive systems(DDS) has been central to the study of non-equilibrium systems in the recent years as they exhibit several interesting features like phase transitions even in one dimension \cite{book2} and have a wide variety of applications \cite{book3} in different branches of physics. A typical example of DDS that has been widely studied in context of both non-equilibrium phase transitions \cite{asep_1} and current \cite{asep_2,asep_3} in NES, is the asymmetric exclusion process \cite{book4}. In particular, asymmetric exclusion process has been used to study extensively current, current fluctuations and its higher order cumulants in \cite{asep_4,asep_5, asep_6} and significant comparison of large deviation functions of current with the notion of equilibrium free energy \cite{asep_7} has been made. Also, the asymmetric exclusion process has also been generalized 
to multi-species \cite{asep_8}, restricted exclusion process \cite{asep_9} and others in different contexts.

In this article we introduce assisted exchange models  on a one dimensional periodic lattice  where
each site can occupy  at most  one particle of type $k=0,1, \dots  K.$   A particle at  any site 
can  exchange their position with  one of  their  nearest neighbors  with a rate  that depends on  both, the type  of particle  pair which are exchanged, and the  type of  particle present at the  left most neighbour 
of the exchanging-pair. We  primarily  address two questions about  the  ($K+1$)- species assisted exchange models (AEM). The first one is to find the exact steady state measure of this non-equilibrium system- in particular  we derive    the  condition  on exchange rates that give rise to pair factorized steady states (PFSS).  We argue that for any finite $K,$ a  pair factorized  steady state  can not  give  rise to  phase separation transition; in other words the systems in this case  remains in a  mixed  phase exhibiting  nontrivial   spatial correlations. We  also aim at obtaining the  exact  steady state current  of each particle species.  It turns out that   AEM    exhibits  density dependent current reversal and negative differential mobility of particle current, which  have been subjects of interest in recent times \cite{crev_azrp, ndm_4}.

Multi-species models   with simple  exchange  dynamics, where  exchange of  different type of  particle pairs  occur with different rates,  have been  introduced earlier \cite{Isaev}.  It turns  out  that steady state  of these  models can not be written in pair-factorized form,  but there can be a matrix product steady state with matrices  satisfying a {\it diffusion algebra}. Some  explicit  examples  of  these  models  with $K=1,2$ have been discussed   in  Refs. \cite{ZRP_rev, Isaev}.  In fact, asymmetric  or symmetric    exclusion process ($K=1$) \cite{book4}  belong to these  class   of models  with $K=1.$  Some other  examples,  with $K=2$   are two species  exclusion models  \cite{ZRP_rev},  ABC model \cite{ABC1, ABC2} and  extended  AHR model \cite{AHR};  some of these models, like ABC  model,  exhibits phase separation transition in one dimension.  

The article is organized  as follows. In section \ref{sec:Model} we introduce the model  and 
derive  a   condition   on  rate functions  that  give rise to a pair-factorized   steady states for a broad 
class of dynamics. Here  we  introduce  a transfer matrix formulation   to  calculate 
the steady state current and spatial correlations. In section III,  we introduce a class of exactly solvable 
assisted exchange  model (AEM) for which the spatial correlations,  particle current and density-fugacity relations   can be   calculated exactly for 
any arbitrary $K.$   Some   specific examples   with $K=1$ and $K=2$ are  discussed in section IV and V respectively. 
We find that  the exchange models  exhibits  several interesting phenomena: reversal of current as  particle density is changed,  negative  differential mobility, double extrema in current voltage relation etc.  Lastly, we conclude in section  VI with some discussions.

\section{Model \label{sec:Model}}
Consider a system of $(K+1)$species of particles on a one dimensional periodic lattice with $L$ sites represented by $i=1,2,\dots,L.$ 
Each site $i$ can be occupied   by  at most one particle of type $k=0,1,\dots,K$; accordingly  the  site variable 
$s_i$  takes  an   non-negative integer value smaller than $(K+1).$  The dynamics of the model  is given by 
\be
  XIJ \mathop{\leftrightharpoons}_{w(XJI)}^{w(XIJ)} XJI, 
  \label{eq:exchange}
\ee
where  $w(XIJ)>0$  are  the exchange rates, which need not have a functional form as the arguments  of $w(.)$  only 
specify the  type of particles. Although $k$   being an  integer   provide  a notion of ordering, the ordering  here does not carry any physical  meaning; in other words the  type particle could have been represented by any  other symbols,  not  necessarily by an integer, like $k=a,b,\dots.$  Clearly   this dynamics  (\ref{eq:exchange}) conserves  the particle number $N_k$  of each kind - the  model has $K$ conservation laws along with  the trivial one  $\sum_{k=0}^K  N_k=L.$

In some examples we discuss here,  $k=0$   are  considered as vacant sites; in that  case  there are  only  $K$-species  of particles of type $k=1,2,\dots,K;$ the exchange  of a particle  with  $0$ present in the  left (or right) neighbor  will  then represent  hopping of that particle  to left  or right - the density conservation of each species   $\rho_k= \frac{N_k}{L}$, where $N_k$  is the  number of particles  of type $k$,   remains unaltered.

\subsection{Steady state}
The Master Equation describing the time evolution of the probabilities of different configurations following the dynamics (\ref{eq:exchange}) 
is as follows
\bea
 \frac{d}{dt}P(\{s_i\})&=&\sum_{i=1}^{L} u(s_{i-1},s_{i+1},s_i)P(\dots s_{i-1},s_{i+1},s_i,s_{i+2}\dots) \cr
                       & -&\sum_{i=1}^{L} u(s_{i-1},s_i,s_{i+1})P(\dots s_{i-1},s_i,s_{i+1},s_{i+2}\dots)
\label{eq:ME}
\eea
Our first goal is to find the steady state by setting the left hand side of Eq.(\ref{eq:ME}) equal to zero. It seems to be quite complex 
to obtain the steady state for the general dynamics (\ref{eq:exchange}) . Instead we look for class of rates for which we can have a pair factorized steady state (PFSS) for 
this assisted hopping and exchange model. In case of PFSS (which by construction is a spatially correlated state in contrast to factorized steady states that may be simpler to achieve but do not contain spatial correlations among its constituents), the steady state weight of 
any possible configuration is expressed as a product of pairs of a function of successive neighbors on the lattice. Explicitly, the steady state weight of any configuration $\{s_i\}$ is given by
\be
 P(\{s_i\})\sim \prod_{i=1}^{L} g(s_i,s_{i+1}) \prod_{k=1}^{K}\delta\left(\sum_{i=1}^{L} \delta_{s_i,k}-N_k\right)
 \label{eq:def_PFSS}
\ee

The right hand side of (\ref{eq:ME}) contains  the  sum of numerous in-flux and out-flux 
terms, which must  equal to zero in   the steady state;  this  cancellation  can  happen  in several ways.  To achieve the PFSS described in 
(\ref{eq:def_PFSS}), it  is sufficient to  follow the condition 
{\small
\bea
&&u(s_{i-1},s_{i+1},s_i)\frac{g(s_{i-1},s_{i+1})g(s_{i+1},s_i)g(s_i,s_{i+2})}{g(s_{i-1},s_{i})g(s_{i},s_{i+1})g(s_{i+1},s_{i+2})} 
- u(s_{i-1},s_i,s_{i+1}) \cr &&~~~~~=h(s_{i-1},s_i,s_{i+1})-h(s_i,s_{i+1},s_{i+2})
\label{eq:cancel_scheme}
\eea
}
where the function $h(.)$ are yet to be determined suitably. Note the right hand side of the condition Eq. (\ref{eq:cancel_scheme}), when summed over all lattice sites gives zero and  ensures 
stationary,  $\frac{d}{dt}P(\{s_i\})=0.$  It is now straight forward to  check that  the follwing 
rate functions  satisfy Eq. (\ref{eq:cancel_scheme}) 
\bea
u(s_{i-1},s_i,s_{i+1})&=&\frac{g(s_{i-1},s_{i+1})}{g(s_{i-1},s_{i})g(s_{i},s_{i+1})}~~~~\mathrm{if}~~~s_{i}\neq s_{i+1} \cr
                      &=& 0 ~~~~\mathrm{if}~~~s_{i}=s_{i+1}    \label{eq:con1}                 
  \eea
with
\bea h(s_{i-1},s_i,s_{i+1})&=&-u(s_{i-1},s_i,s_{i+1}) ~~~~\mathrm{if}~~~s_{i}\neq s_{i+1} \cr                                           &=& 0 ~~~~\mathrm{if}~~~s_{i}=s_{i+1}.
\eea


\subsection{Transfer matrix formulation}
The next task would be to calculate the partition function and  the   observables  of the $(K+1)$-species  exchange model, which is, 
\bea
Q(\{N_k\}) &=&  \sum _{\{s_i\}}\prod_i  g(s_i,s_{i+1}) 
\prod_k \delta \left(  \sum_i \delta_{s_i,k} - N_k \right).  
\eea
The  $\delta$-functions   ensure that  the particle numbers  ($N_k$-s)   are conserved.
We now work  in  the grand canonical ensemble(GCE) 
and  associate fugacities $\{ z_k\},$  one for  each species, which will control  the  particle densities $\{ \rho_k\}.$  Also we set $z_0=1,$  without loss of  generality. 
 Hence the partition function in the GCE  is 
 \bea
Z(\{z_k\}) &=& \sum_{\{N_k\}}^\infty  Q(\{N_k\}) \prod_k (z_k)^{N_k}\\
&=&  \sum _{\{s_i\}}\prod_i  z_{s_i}g(s_i,s_{i+1}) =\mathrm{Tr}[T^L] 
\eea
where  $T$ is a 
$(K+1)$ dimensional square  matrix
\be
T= \sum_{k',k=0}^K g(k',k)z_{k} |k'\rangle\langle k|
\label{eq:tm_gen}
\ee
which is  formally known as the transfer matrix. Here $\{ |k \rangle\}$  with $k=0,1,\dots,K$ are  
the  standard   basis  vectors in $(K+1)$-dimension.  The transfer matrix  $T$ can  also be written as 
\be
T = \sum_{k=0}^K  z_k D_k ~~ {\rm with}~~ D_k = \sum_{k'=0}^K g(k',k)|k'\rangle\langle k| \label{eq:mpa_rep}
\ee

where the matrix $D_k$ represents a particle of the species $k.$  With these set  of matrices  $\{ D_k \}$  we  write the steady state  weights  of the system in matrix product form
\be 
P(\left\lbrace s_i \right\rbrace)   \sim   \prod_i g(s_i, s_{i+1})  = \mathrm{Tr} [\prod_i D_{s_i}] \label{eq:MPAform}
\ee
In matrix product form,  the  correspondence  of particles  by a  representing matrix,  helps 
 in calculating expectation values of several observables, which is discussed below.

With Eq.(\ref{eq:mpa_rep}) in hand, we can proceed to calculate different observables analytically. Let us start with density $\rho_k$ of the particles of species $k.$ 
\bea
\rho_k=\la k\ra =\frac{\mathrm{Tr}[z_k D_k T^{L-1}]}{\mathrm{Tr}[T^L]}=\frac{\langle k|T^L|k\rangle}{\mathrm{Tr}[T^L]},
\label{eq:rho1}
\eea
Let the eigenvalues of $T$ are $\lambda> \lambda_1\ge \lambda_2 \dots \ge \lambda_K$ with corresponding 
right and left eigenvectors (normalized)  $\{|\psi\ra, |\psi_1\ra,|\psi_2\ra \dots |\psi_K\ra\}$ and 
 $\{\la \phi|, \la\phi_1|,\la\phi_2| \dots \la \phi_K|\ra\}$  respectively.  Since  $T$ is a positive 
 matrix (as   $g(i,j)>0),$ the largest  eigenvalue $\lambda$  is  non-degenerate  and  the corresponding eigenvector $|\psi\ra$  can be chosen positive.    
Thus, 
\bea 
T^n = \lambda^n |\psi\ra \la \phi|  + \sum_{k=1}^K \lambda_k^n |\psi_k\ra \la \phi_k|
\eea
Using this in Eq. (\ref{eq:rho1})  we calculate the density  of each species   in  the thermodynamic limit  $L\rightarrow \infty$  ( i.e. $\lambda^L \gg \lambda_k^L ~ \forall 1\le k\le K$),
\bea
\rho_k = \la k|\psi\ra\la \phi|k\ra. \label{eq:density}
\eea
In a similar way one can calculate expectation values, that species  $k$ is a   right neighbor of $m$ 
in steady state, 
\bea
\la mk \ra &=&\frac{\mathrm{Tr}[z_mz_k D_m D_k T^{L-2}]}{\mathrm{Tr}[T^L]}\cr
    &=&  \frac{z_k g(m,k)}{\lambda} \la k|\psi\ra\la\phi|m\ra,
\eea
where in the last step we have used the thermodynamic limit.
In a similar manner, one can obtain the 
two-point spatial correlation functions between any two species of particles (say $k$ and $k'$) at a distance $r$,
\bea
C_{k,k'}(r)&=&\frac{\mathrm{Tr}[z_k D_k T^{r-1}z_{k'} D_{k'} T^{L-r-1} ]}{\mathrm{Tr}[T^L]}-\rho_k \rho_{k'} \cr
&=&\frac{\langle k|T^r|k'\rangle\langle k'|T^{L-r}|k\rangle}{\mathrm{Tr}[T^L]}-\rho_k \rho_{k'}\cr
&=&  \langle k'| \psi \ra \la \phi|k\rangle  \sum_{m=1}^K \langle k|\psi_m \ra \la \phi_m|k'\rangle 
\left(\frac{\lambda_m}{\lambda}\right)^r.
\label{eq:corr_gen}
\eea
Here,  to obtain the last  step, we used Eq. (\ref{eq:density})  and taken  the thermodynamic limit.  For large $r,$ the dominant contribution   to the  correlation function comes from the  first term $m=1$ (in the sum), i.e. 
$C_{k,k'}(r)\sim ~  (\frac{\lambda_1}{\lambda})^r =  e^{-r/\xi}$ where   the correlation length $\xi 
= (\ln \frac{\lambda_1}{\lambda})^{-1}.$  

From the  study  of correlation functions it is clear that  a pair factorized steady state  can not give rise to 
diverging correlation length  if  the number of species  are  finite.  For  $(K+1)$ species model, one gets 
$(K+1)-$ dimensional transfer matrix  with  elements  $\la m |T|n\ra= z^m g(m,n)>0;$  the largest  eigenvalue 
$\lambda$ then remains  non-degenerate  following Perron-Frobenius theorem. Thus, the correlation 
length $\xi = (\ln \frac{\lambda_1}{\lambda})^{-1}$   is  finite and    possibility of  phase transition 
is ruled out  in   $(K+1)$ component systems  with  PFSS. 
 To get  out of  this  situation, one   may think of  setting  some matrix elements to zero  so  that 
the  transfer matrix   become reducible and  then, Perron-Frobenius theorem (non-degenerate largest  eigenvalue) 
does not apply. However,  it is easy to check  that a reducible form of $g(m,n)$ force the  steady state weight of 
all configurations   to be zero.

There  are numerous examples of  exchange models   which exhibit  phase transition; like extended KLS 
models  with ferro \cite{EKLS} or anti-ferromagnetic \cite{AF} interactions,  ABC models  in an interval 
\cite{ABC2} or a ring \cite{ABC1}.  However, the steady state  of these  models  are not  pair factorized. 
In some  models,   the steady state  can be written in matrix product form \cite{MPA}  with matrices 
having dimensions  larger than the  number  of  components $(K+1)$,  then  the matrix  elements of 
corresponding transfer matrix  can not be treated as  the  weight factors, as   in PFSS. In this case 
one   can have a reducible transfer matrix which does not restrict the phase space  even though a  block 
of matrix elements  are  zero.

The assisted  exchange  models, with factorized steady state,  does not   give rise to phase separation transitions in general, 
but they  exhibit  interesting  steady-state  current behavior which will be discussed in the next section.

\subsection{Average current of particles}
The  average current is an entity of interest since the non-equilibrium phenomena are characterized by a net flow or current in the system comparing to their equilibrium counterparts which are dictated by zero average current. In this section we will focus on calculating exactly the average particle current in the pair factorized steady states of $(K+1)$species assisted  exchange models. In particular, the average current for the species $k$ would be

\be
\langle J_k\rangle= \sum_{k'\ne k}^K \sum_{m=0}^{K}u(m,k,k')\langle m k k'\rangle-u(m,k',k)\langle m k' k\rangle
\label{eq:current_gen}
\ee
Using the matrix representations in Eq. (\ref{eq:mpa_rep}), it is quite straightforward to obtain 
\bea
\langle m k k'\rangle&=&\frac{z_k z_{k'}}{\lambda^2} g(m,k) g(k,k') \la k'|\psi\ra \la  \phi|m\ra\cr
  &=&\frac{ \la mk\ra \la kk'\ra}{\rho_k}
\eea
Then, for the exchange dynamics (\ref{eq:con1}), the current of species $k$ is 
\bea
\langle J_k\rangle&=&\frac{1}{\lambda^2} \sum_{k'\ne k}^K \sum_{m=0}^{K} z_k z_{k'} \la\phi|m\ra 
\left(  g(m,k')  \la k'|\psi\ra - g(m,k) \la k|\psi\ra \right)\cr
&=& \frac{1}{\lambda}\sum_{k'\ne k}^K \sum_{m=0}^{K}\left( z_k \la mk'\ra - z_{k'} \la mk\ra \right)
\label{eq:cur_cond1}
\eea

To proceed  further we need to be more specific about the dynamics.   
In the  following section we choose a 
specific  form of  weight function $g(m,n)$ for which the steady state 
results for  current and other observables can be obtained  exactly  for 
any  any arbitrary $K.$

\section{Exact results for a class of AEM }
In this section  we choose a specific  form of  weight function $g(m,n)$ for which the steady state calculations of 
current can be done explicitly for any arbitrary $K.$  Let us consider   a  weight function 
\bea
g(m,n)&=&\frac{g(m,0)g(0,n)}{\gamma g(0,0)} ~~~m,n>0,
\label{eq:con2_g}
\eea
where $g(m,0)$ and $g(0,n)$  are $2K + 1$ independent  parameters; once these parameters are fixed, the rest of the  elements  of $g(m,n)$ can be calculated using  Eq. (\ref{eq:con2_g}).
In the following, for convenience, we set a  short-hand  notation
\bea
v_m = g(m,0) g(0,m).
\eea
It is easy to check that the transfer matrix  $T,$ with elements $\la m |T|n\ra = z_m g(m,n)$ along with Eq. (\ref{eq:con2_g}),  has the following properties, 
\bea
&I.& ~ \mathrm{Tr}[T] =\sum_{k=0}^K z_k g(k,k)\equiv 2 \tau \cr 
&II.& ~\mathrm{Det}[T] = 0 \cr
&III.& ~{\rm Eigenvalues:}  
\left\{ \lambda_{\pm}=  \tau \pm \sqrt{\tau^2 - \delta}, 0,0,\dots \right\}, ~~~~\delta \equiv (\gamma^{-1}-1)\prod_{k=1}^{K}z_k g(k,k)
\eea
Let $|\psi \ra,$  and $\la \phi|$ be the left  and right eigenvectors (normalized), corresponding to the  largest eigenvalue $\lambda\equiv \lambda_+$  with elements,   $\la m|\psi \ra,$  and $\la \phi|m\ra$ with  $m=0,1,2, \dots K.$   For  $m>0$ we have, 
\bea
\la m|\psi \ra &=& \eta ~g(m,0) \la 0|\psi \ra; ~ \la \phi|m\ra = \eta~ z_m g(0,m)\la \phi|0\ra; ~~{\rm with} ~~\eta= \frac{\lambda +(\gamma-1) g(0,0)}{\gamma g(0,0)\lambda}.
\eea
For $m=0,$   one  can determine  $\la 0|\psi \ra$ and $\la \phi|0\ra$ from  the normalization  condition  
$\la \phi|\psi \ra=1,$
\bea
\la \phi|0\ra\la 0|\psi\ra = \left( 1+ 2 \eta^2 \tau -\eta^2 g(0,0)\right)^{-1}
\eea

Using these properties of  $T$ it is straight forward to calculate  the  observables - we  state  the results 
in the following.  Let us  use the grand canonical ensemble 
and  remind  ourselves that,  without loss of generality,  we  can set  the fugacity  
$z_0=1.$  The particle densities are then 
\bea
\rho_m =\frac{\gamma z_m v_m  g^2(1,1)}{2 \alpha'\alpha} \left[ \alpha' + \alpha -v_1^2\right]
\eea
where, 
\bea
\alpha'=\sqrt{(\alpha + v_1^2)^2+4(\gamma-1)\alpha v_1^2 }\cr
 {\rm and} ~\alpha=\gamma g^2(1,1)\sum_{k=1}^{K}z_k   v_k.
\eea

Similarly,  the  correlation function    is
\bea
C_{m,m'}(r)=\frac{\gamma z_{m'} g(m,0)g(0,m')g^2(1,1)}{2 \alpha'\alpha} \left[ \alpha' - \alpha +v_1^2\right]
            \left(\frac{\lambda_{-}}{\lambda}\right)^r \nonumber
\eea

This being stated, we now proceed to obtain the particle current given by (\ref{eq:cur_cond1}). From now onwards, we will be concerned 
with the average current of the species $k=0.$ As we have already mentioned that $k=0$ can be considered as vacant sites and $k>0$  as  the  $K$-particle  species;   the  particles $k=1,2,\dots,K$ exchange  with each other whereas the exchange of any species $k$ with $0$ will now represent diffusion, i.e.,  hopping of that species to right or left vacant neighbour.  Clearly in this case the total current including that of $0$ or vacant sites will be $\la J \ra + \la J_0 \ra = 0$ where $\la J \ra$ is the total particle current of  $K$-species of particles $k=1,2,\dots,K.$ Then, 
$\la J \ra=-\la J_0 \ra.$  Current $\la J_0 \ra$ can be calculated  from Eq. (\ref{eq:cur_cond1}), which gives  the total  particle current 
\bea
&&\langle J\rangle=\frac{1}{\lambda^2}\left[(\rho_0 g(0,0)+\sqrt{\rho_0}s_k)\sum_{i=1}^{k}z_i 
                                                - (\sqrt{\rho_0}+\frac{1}{\gamma g(0,0)}s_k)s_k\right] \label{eq:cur_cond1_g_cons} \\
 &&\mathrm{where}~~~s_k=\sum_{i=1}^{k} \sqrt{\rho_i z_i v_i}.\nonumber
\eea
Inverting the density fugacity relations, we arrive at the following equation
\bea
 \hspace*{-2.0cm}z_i=\frac{\rho_i}{v_i}\frac{(v_1)^{K-1}}{2g^2(1,1)\rho_0(1-\rho_0)^2}
 [~~v_1(1-2\rho_0)^2+2g(0,0)g(1,1)\rho_0(1-\rho_0) \cr
    +(1-2\rho_0)\sqrt{v_1(v_1(1-2\rho_0)^2+4g(0,0)g(1,1)\rho_0(1-\rho_0))}~~],\cr
\label{eq:fugacity_den}
\eea
that expresses the fugacities $z_i(i=1,2,\dots,k)$ in terms of the densities $\rho_i$ so that $z_i$ can be replaced in Eq. (\ref{eq:cur_cond1_g_cons}). Note the difference between the expressions of the average particle current $\langle J\rangle$ 
in (\ref{eq:cur_cond1_g_cons}) compare to (\ref{eq:cur_cond1})-in case of  (\ref{eq:cur_cond1_g_cons}), it became possible to write down 
the current only in terms of the input parameters like the densities and hop rates(after replacing $z_i$s using (\ref{eq:fugacity_den}))
-which was not that straightforward for (\ref{eq:cur_cond1}).

Now, as we have obtained the exact average current for  the $K-$species assisted exchange model studied in 
this article, our next move would be to investigate possible interesting properties of the current. First we  consider next a 
simple case with $K=1$ i.e. a single species of particles undergoing assisted hopping on a $1-$D periodic lattice and we will discuss 
phenomenas like density dependent current reversal, negative differential mobility of particles in detail for this simple example. 

\section{Assisted exchange  model with $K=1$ }
Consider a $1-$D periodic lattice with $L$ sites $i=1,2,\dots,L$ where each site $i$ can be occupied by at most one particle({\it i.e}.the particles are hard core), each represented by $1$ or it can be vacant, represented by $0$. A particle from a randomly chosen site $i$ 
can move to its right neighbor(if vacant) $(i+1)$ with a rate that depends on the left neighbor $(i-1)$ of the departure site. Whereas, 
the particle from $i$ can also move to its left neighbor(if vacant) $(i-1)$ with a rate that depends on the left neighbor $(i-2)$ of the 
arrival site $(i-1)$. More precisely, in a nutshell, the motion of the particles are assisted by their neighbors- isolated particles and 
crowded particles(particles with occupied neighbors) hop in different manner. The dynamics is represented as
\be
010 \mathop{\rightleftharpoons}_{q}^{p} 001~~~~~~~~~~110 \mathop{\rightleftharpoons}_{\gamma_2 q }^{\gamma_1 p} 101
\label{eq:example_k_1}
\ee

For this  dynamics  we  have  pair factorized steady state,  following (\ref{eq:con1}) 
only if $\gamma_1= \frac{1}{\gamma_2} \equiv \gamma$; the  pair-weight functions are  then

\bea
g(0,0) = \frac1q;~  g(1,0) g(0,1)=v_1= \frac{1}{pq}; g(1,1) = \frac{1}{\gamma p}\label{eq:con1_con2_same}
\eea
and the  corresponding dynamics, 
\be
010 \mathop{\rightleftharpoons}_{q}^{p} 001~~~~~~~~~~110 \mathop{\rightleftharpoons}_{ q\gamma^{-1} }^{\gamma p} 101
\label{eq:example_k_1a}
\ee
has three    parameters  $p,q$ and $\gamma.$
It is easy to show that the particle current for this single species assisted exchange model ultimately becomes 
\bea
\langle J\rangle
                =p (\langle 010 \rangle +\gamma \langle 110 \rangle)-q(\langle 001 \rangle +\frac{1}{\gamma} \langle 101 \rangle)
                =\frac{1}{\lambda^2}\left[z_1(\frac{\rho_0}{q}-\frac{\rho_1}{\gamma p})+(z_1-1)\sqrt{\frac{\rho_0\rho_1z_1}{pq}}\right] \label{eq:cur_example_k_1} \\             
\mathrm{where}~~~\lambda=\frac{1}{2}\left(\frac{1}{q}+\frac{z_1}{\gamma p} +\sqrt{(\frac{1}{q}-\frac{z_1}{\gamma p})^2+4 \frac{z_1}{pq}}\right).\nonumber
\eea
In the above equation, we have calculated the current in the grand canonical ensemble by associating a fugacity $z_1$ to the particles. In 
order to express the average current $\langle J\rangle$ in terms of particle or vacancy densities, one can replace $z_1$ in 
(\ref{eq:cur_example_k_1}) by inverting the density-fugacity relation as follows 
\bea
z_1&=&\frac{\gamma p}{q}+\frac{(1-2\rho_0)\gamma^2 p^2}{2 \rho_0(1-\rho_0)} 
\left[\frac{1}{p q}(1-2\rho_0)+\sqrt{\frac{1}{p q}(\frac{(1-2\rho_0)^2}{p q}+4\frac{\rho_0(1-\rho_0)}{\gamma p q})}\right].~~~~~
\label{eq:fugacity_example_1}
\eea
Now we would like to discuss two interesting features viz. density dependent current reversal and negative differential mobility of 
particles emerging from the expression of particle current in Eq. (\ref{eq:cur_example_k_1}) that resulted from the stochastic process 
(\ref{eq:example_k_1}) along with (\ref{eq:con1_con2_same}).

\subsection{Current reversal in AEM  with $K=1$}
Let us discuss a more specific example with, $q= \frac1p$ and $\gamma = \frac{1}{p^2};$ the dynamics
is then,  
\be
010 \mathop{\rightleftharpoons}_{\frac{1}{p}}^{p} 001~~~~~~~~~~110 \mathop{\rightleftharpoons}_{p}^{\frac{1}{p}} 101~~.
\label{eq:example_k_1_simplified}
\ee
Now comparing the rates in Eq. (\ref{eq:example_k_1_simplified}) with that in (\ref{eq:example_k_1}) and further using 
Eq. (\ref{eq:con1_con2_same}), we conclude that the dynamics in (\ref{eq:example_k_1_simplified}) indeed gives rise to a pair factorized 
steady state with the pair weight factors obtained as $g(1,0)=g(0,1)=1$ and $g(0,0)=g(1,1)=p$.
Correspondingly, the particle current from Eq. (\ref{eq:cur_example_k_1}) will be 
\be
\langle J\rangle=\frac{1}{\lambda^2}\left[z_1 p(\rho_0 -\rho_1 )+(z_1-1)\sqrt{\rho_0\rho_1z_1}\right],
\ee
where $\lambda=\frac{1}{2}(p(1+z_1)+\sqrt{p^2(1-z_1)^2+4z1})$. But the above expression can be further simplified if 
we use the fact $\rho_0=1-\rho_1$ and also note that the fugacity in Eq. (\ref{eq:fugacity_example_1})  subsequently becomes
\be
z_1=1+\frac{(2\rho_1-1)}{2p^2\rho_1(1-\rho_1)}\left[2\rho_1-1+\sqrt{(2\rho_1-1)^2+4p^2\rho_1(1-\rho_1)}\right].
\label{eq:fugacity_density_k_1}
\ee
Then the current takes the functional form as follows
\bea
\hspace*{-2.0cm}\langle J\rangle=\frac{(2\rho_1-1)}{\lambda^2}\left[p z_1 +\frac{\sqrt{\rho_1(1-\rho_1)z_1}}
{2p^2\rho_1(1-\rho_1)}\left(2\rho_1-1+\sqrt{(2\rho_1-1)^2+4p^2\rho_1(1-\rho_1)}\right)\right]~~~~~
\label{eq:cur_example_k_1_simplified}
\eea
Note that in principle, we should have replaced $z_1$ all the way through Eq. (\ref{eq:cur_example_k_1_simplified}) by 
using (\ref{eq:fugacity_density_k_1}), but keeping $z_1$ intact in some places other than the required does not harm 
our purpose of showing density dependent current reversal from (\ref{eq:cur_example_k_1_simplified}). More precisely, 
to observe current reversal by changing particle density for a fixed dynamics, we have to have the current 
$\langle J\rangle$ in Eq. (\ref{eq:cur_example_k_1_simplified}) equal to zero for some non-zero value of the particle 
density $\rho_1=\rho^\ast_1.$ As a careful observation ensures the fact that the entities involving $z_1$ in the expression 
of $\langle J\rangle$ in Eq. (\ref{eq:cur_example_k_1_simplified}) are always greater than zero, so the only possibility 
to obtain density dependent current reversal i.e. $\langle J\rangle=0$ in (\ref{eq:cur_example_k_1_simplified}) for 
the fixed dynamics (\ref{eq:example_k_1_simplified}) is to set $\rho^\ast_1=\frac{1}{2}.$

\begin{figure}
\vspace*{.5 cm}
 \centering
\includegraphics[height=8.5 cm]{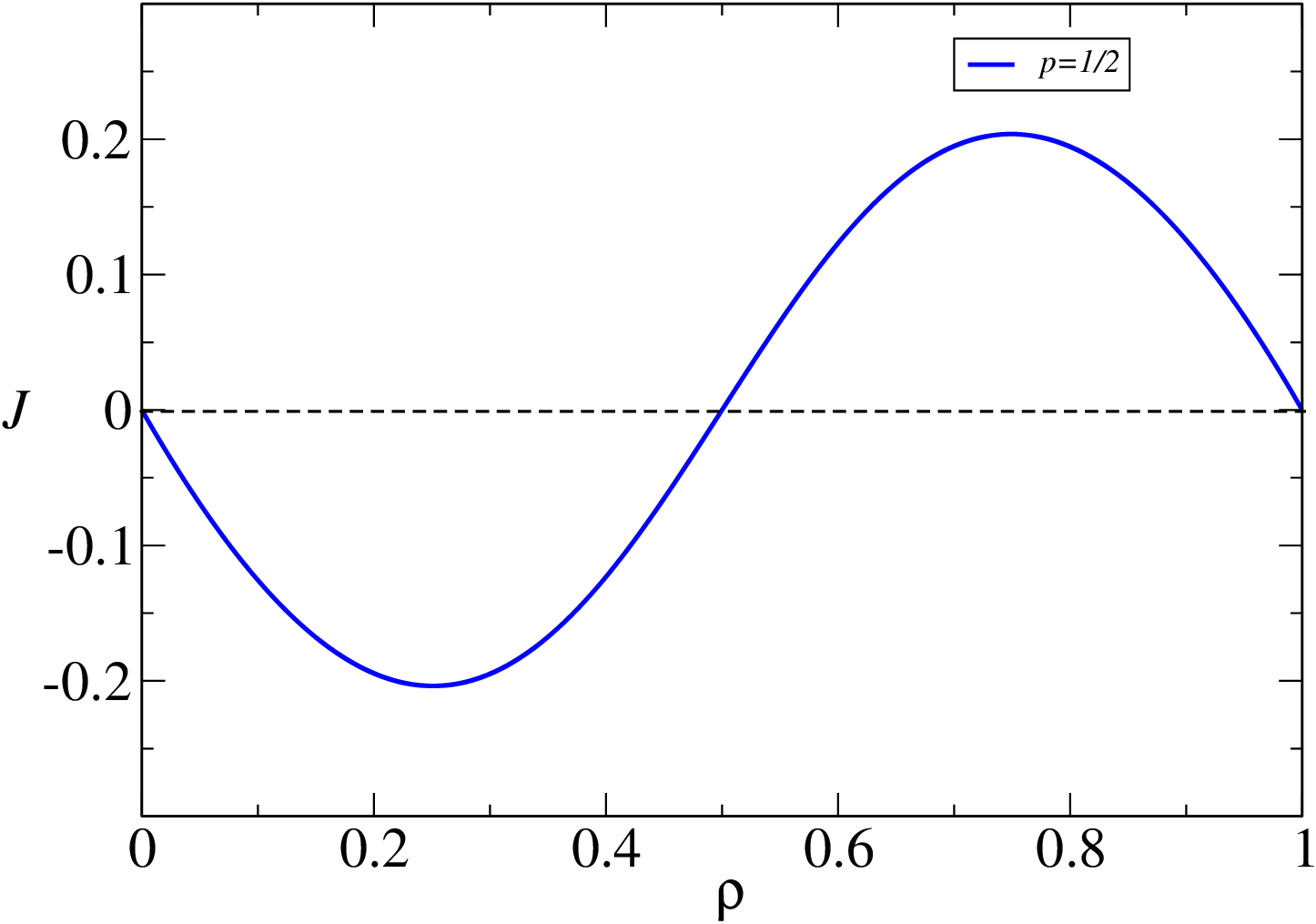}
 \caption{Current as a function of density for $K=1.$ Density dependent current reversal for single species assisted hopping model obeying dynamics (\ref{eq:example_k_1_simplified}) 
 with $p=\frac{1}{2}$- the current becomes zero at $\rho^\ast=\frac{1}{2}$(point of reversal) and then reverses its direction as density is 
 further increased.}
 \label{fig:1}
\end{figure}
To summarize, for the given dynamics (\ref{eq:example_k_1_simplified}) with some fixed value of $p,$ if we start with 
a very small particle density $\rho_1\approx 0$ and increase the density gradually, at first the average current 
flows towards right(if $p>1$) starting from zero(or, towards left(if $p<1$)), reaches a maximum(or, minimum) value, then 
decreases(or, increases) continuously until at some particular value of density $\rho^\ast_1=\frac{1}{2}$- which we call 
to be the {\it point of reversal}- the current $\langle J\rangle$ becomes zero, after which if we increase the particle 
density, current starts flowing in the opposite direction namely towards left (or, towards right). This incident of change 
in the direction of particle current as we tune the density for a fixed dynamics, is known as density dependent current 
reversal and that particular value of density at which the current equals to zero just before changing its direction, is 
termed as the point of reversal. We can see in Fig. \ref{fig:1} that the particles following the dynamics (\ref{eq:example_k_1_simplified}) with $p=\frac{1}{2},$ undergo a reversal in the direction of current with $\rho^\ast_1=\frac{1}{2}$.

\subsection{Negative differential mobility in AEM  with $K=1$}
In this section, we consider yet another specific case, where  $p=1, q= e^{-\epsilon}$ 
\be
010 \mathop{\rightleftharpoons}_{e^{-\epsilon}}^{1} 001~~~~~~~~~~
110 \mathop{\rightleftharpoons}_{e^{-\epsilon}\gamma^{-1}}^{\gamma} 101~~; \gamma= \frac{1}{1+\epsilon/2}
\label{eq:example_k_1_negr}
\ee
The above hop rates, when compared to Eq. (\ref{eq:example_k_1}) along with (\ref{eq:con1_con2_same}), imply that dynamics 
(\ref{eq:example_k_1_negr}) lead to a pair factorized steady state with 
$g(1,0)=g(0,1)=e^{\epsilon/2}$,$g(0,0)=e^{\epsilon}$ and $g(1,1)=(1+\epsilon/2)$.  
Note that, in absence of the additional factor $\epsilon/2$ in (\ref{eq:example_k_1_negr}) the isolated particles 
and crowded particles hop with same rates and the model no longer remains an assisted one. As usual, we define the net bias 
on the particles towards some direction, say right, as the logarithm of the ratio of the right hop rate to the left one. 
More elaborately, the bias on the isolated particles is $\mathrm{ln}(\frac{1}{e^{-\epsilon}})$ i.e. simply $\epsilon$ whereas 
the crowded particle feels a rightward bias given by $(\epsilon-2\mathrm{ln}(1+\epsilon/2)).$ Since, both the bias are 
monotonically increasing functions of $\epsilon,$ from here on, we simply consider $\epsilon$ as the bias and will discuss 
the behavior of the particle current for (\ref{eq:example_k_1_negr}) as we change the bias $\epsilon$ for a given particle 
density. Firstly, from Eq. (\ref{eq:cur_example_k_1}), we calculate the average current $\langle J\rangle$ for the present dynamics 
(\ref{eq:example_k_1_negr}),
\be
\langle J\rangle=\frac{1}{\lambda^2}\left[z_1(\rho_0 e^{\epsilon}-\rho_1(1+\epsilon/2))+(z_1-1)\sqrt{\rho_0\rho_1z_1e^{\epsilon}}\right]
\label{eq:cur_negr_k_1}
\ee
where $\lambda=\frac{1}{2}\left(e^{\epsilon}+z_1(1+\epsilon/2) +\sqrt{(e^{\epsilon}-z_1(1+\epsilon/2))^2+4z_1e^{\epsilon}}\right)$ 
and from the density-fugacity relation, one can also express the fugacity $z_1$ in terms of the densities and other parameters as follows 
\bea
\hspace*{-1.0cm}z_1=\frac{e^{\epsilon}}{(1+\epsilon/2)}+\frac{(1-2\rho_0)}{2(1+\epsilon/2)^2 \rho_0(1-\rho_0)}\times\cr
\left[e^{\epsilon}(1-2\rho_0)+\sqrt{e^{\epsilon}(e^{\epsilon}(1-2\rho_0)^2+4e^{\epsilon}(1+\epsilon/2)\rho_0(1-\rho_0))}\right].
\eea
Let us be more specific and analyze the system at density $\rho_1=\frac{1}{2}=\rho_0.$ Then the fugacity $z_1$ in the above equation simply 
relates to the bias $\epsilon$ as
\be
z_1=\frac{e^{\epsilon}}{(1+\epsilon/2)}.
\ee
Correspondingly, substituting this value of $z_1$ in Eq. (\ref{eq:cur_negr_k_1}), the particle current at $\rho_1=\frac{1}{2}=\rho_0$ 
becomes
\be
\langle J\rangle=\frac{1-e^{-\epsilon}(1+\frac{\epsilon}{2})}{2(1+\frac{\epsilon}{2}+\sqrt{1+\frac{\epsilon}{2}})}
\label{eq:cur_negr_k_1_density_half}
\ee
\begin{figure}
\vspace*{.5 cm}
 \centering
\includegraphics[height=8.5 cm]{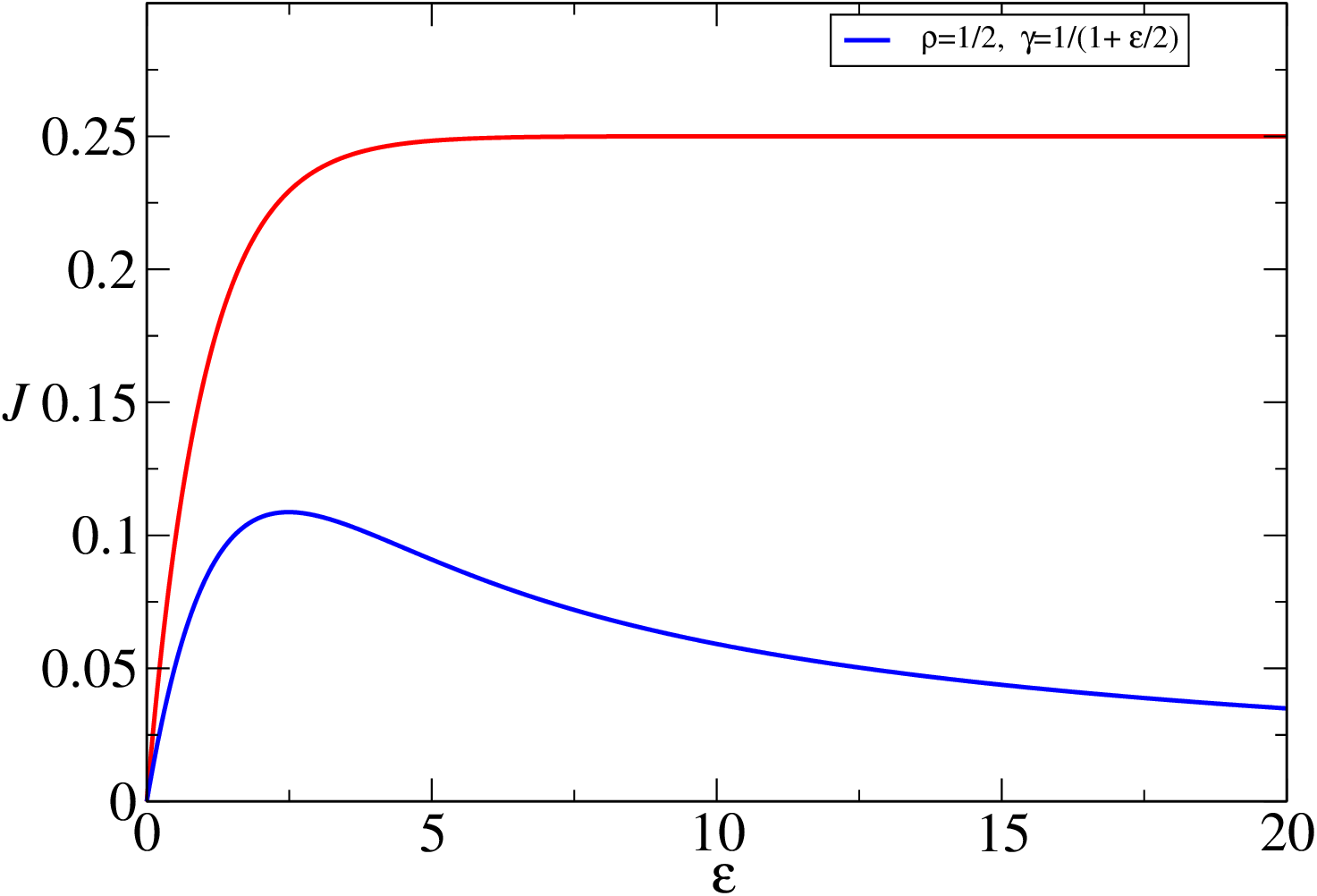}
 \caption{Negative differential mobility exhibited by particles of the single species assisted hopping model when the dynamics 
 (\ref{eq:example_k_1}) takes the specified form of (\ref{eq:example_k_1_negr}) in contrast to another specific case of 
 (\ref{eq:example_k_1})  where $\gamma=1$ - both the systems are studied at particle density $\rho_1=\frac{1}{2}$.}
 \label{fig:2}
\end{figure}
If we plot the current($\langle J\rangle$) in Eq. (\ref{eq:cur_negr_k_1_density_half}) as a function of the bias($\epsilon$), as shown in 
Fig. \ref{fig:2}. by the green curve, we observe 
that the current is zero at $\epsilon=0$ (as expected from dynamics (\ref{eq:example_k_1_negr}) since $\epsilon=0$ leads to the equilibrium 
situation), then as we increase the bias the current increases. But at a finite value  $\epsilon=\epsilon^\star\approxeq 2.4,$ $\langle J\rangle$ 
reaches to its maximum value, after which, {\it as we increase the bias the current decreases gradually giving rise to negative differential 
mobility} (i.e. $\frac{\mathrm{d}\langle J\rangle}{\mathrm{d}\epsilon}<0$ in the parameter region $\epsilon^\star <\epsilon <\infty$). 
This incidence of negative response is evident from the expression (\ref{eq:cur_negr_k_1_density_half}) because the current approaches to 
zero in both the limit $\epsilon\rightarrow0$ and $\epsilon\rightarrow\infty$, so in between $0<\epsilon<\infty,$ there must be finite 
$\epsilon=\epsilon^\star$ at which the current reaches to an extremum and in one of the regions(either greater than or less than 
$\epsilon^\star$), the current as a result exhibits negative differential mobility. To emphasize the role of the factor $\epsilon/2,$ we 
present another curve (the red one) in Fig. \ref{fig:2} that corresponds to the current in absence of the factor $\frac{\epsilon}{2}$ in 
(\ref{eq:example_k_1_negr}), 
more accurately, it corresponds to the dynamics (\ref{eq:example_k_1}) with $p=1,q=e^{-\epsilon},\gamma=1$. It is quite 
straightforward to prove that this particular dynamics results in a factorized steady state and the current there is simply 
$\langle J\rangle=(1-e^{-\epsilon})\rho(1-\rho)$. So, the response for $\rho_1=1/2=\rho_0$ in this case is just 
$\frac{\mathrm{d}\langle J\rangle}{\mathrm{d}\epsilon}=\frac{e^{-\epsilon}}{4}$ which remains positive always irrespective of the value of 
$\epsilon$- this ensures that in absence of the additional factor $\epsilon/2$ the dynamics (\ref{eq:example_k_1_negr}) does not give 
rise to negative differential response and the corresponding steady state current only increases with increasing bias and ultimately, 
for large values of $\epsilon$, more precisely in the limit $\epsilon\rightarrow\infty,$ settles to a finite constant value 
$\rho(1-\rho)=\frac{1}{4}$. 

Let us discuss qualitatively the difference between the situations with $\gamma=1$(red curve in Fig. that does not show negative 
differential mobility) in comparison to the case where $\gamma=1/(1+\epsilon/2)$(blue curve in Fig. exhibiting negative 
differential response in the regime $<\epsilon<\infty$) in dynamics (\ref{eq:example_k_1}). Actually in both the cases, the isolated particles hop with same rates but the crowded particles (meaning when particle hopping is assisted by neighboring particles, $110\rightleftharpoons 101$) differ. 
A  part  of this  dynamics where  the   the immediate neighbour (of the of the 
three  sites   under consideration)  is  vacant, is  
\bea
1100 \mathop{\rightleftharpoons}_{e^{-\epsilon}\gamma^{-1}}^{\gamma} 1010.
\label{eq:example_k_1_compare_4_sites}
\eea

So when a crowded particle moves to right, it creates isolated particles(1's) and destroys 11-pairs causing de-clustering or breaking of cluster of particles(1's) creating more activity- whereas the corresponding left move 
decreases the number of isolated particles and increases number of 11-pairs i.e. the left move causes clustering of particles(1's) implying 
decrease in activity. It is evident from Eq. (\ref{eq:example_k_1_compare_4_sites}) that due to the presence of the additional factor 
$\epsilon/2$ in the later case, as we increase $\epsilon,$ for large drive, the clustering of 1-s through the leftward move 
increase significantly for the dynamics corresponding to $\gamma=1/(1+\epsilon/2)$ in comparison to the case $\gamma=1$. 
Clusterization of 1's with increasing bias in turn implies decrease in activity which results in decrease of particle current as 
$\epsilon$ increases. Consequently we have negative differential mobility of particles for the dynamics (\ref{eq:example_k_1_negr}). 

This concludes our discussion about the single species assisted hopping model with a generic dynamics of the form (\ref{eq:cur_negr_k_1})- 
which possess pair factorized steady state when the rates are specified by the condition (\ref{eq:con1_con2_same}). We have shown 
explicitly that this model, with some specific choice of rates, as given by Eq. (\ref{eq:example_k_1_simplified}) and 
(\ref{eq:example_k_1_negr}) can exhibit density dependent current reversal(Fig. \ref{fig:1}) and 
negative differential mobility(Fig. \ref{fig:2}) respectively. 
%

\section{Assisted exchange  model with $K=2$}
In this section we will  we describe one  example of the dynamics Eq.   for  $K=2.$ 
We start by considering a specific rate function, 
\bea
g(0,0)&=&\frac{3}{10},g(0,1)=\frac{1}{5},g(0,2)=\frac{2}{5},g(1,0)=\frac{1}{2},\cr
g(1,1)&=&\frac{1}{2},g(1,2)=1,g(2,0)=\frac{1}{10},g(2,1)=\frac{1}{8},g(2,2)=\frac{1}{10}.
\eea
For  exchange rates  can be constructed from the pair weight  function $g(.,.)$ in a straightforward way,  using 
Eq. (\ref{eq:con1}). The fugacities $z_1$ and $z_2$ corresponding to the particles 
of the two different species, can now be expressed as a function of the corresponding particle densities $\rho_1,\rho_2$ 
as(using Eq. (\ref{eq:fugacity_den})),
\bea
\hspace*{-2.5cm}z_1(\rho_1,\rho_2)= \frac{2\rho_1}{\rho_0(1-\rho_0)^2}\left[(1-2\rho_0)^2
+3\rho_0(1-\rho_0)+(1-2\rho_0)\sqrt{(1-2\rho_0)^2+6\rho_0(1-\rho_0)}\right] \cr
\hspace*{-2.5cm} z_2(\rho_1,\rho_2)= \frac{\rho_2}{2\rho_0(1-\rho_0)^2}\left[(1-2\rho_0)^2
+3\rho_0(1-\rho_0)+(1-2\rho_0)\sqrt{(1-2\rho_0)^2+6\rho_0(1-\rho_0)}\right]~~~~~
\label{eq:fug_den_k_2}
\eea
and the particle current, as given by (\ref{eq:cur_cond1_g_cons}), takes the following form
\bea
\langle J\rangle(\rho_1,\rho_2)=\frac{1}{\lambda^2}\left[\left(\frac{3}{10}\rho_0 +\sqrt{\rho_0}s_2\right)(z_1+z_2)
                                                - (\sqrt{\rho_0}+5s_2)s_2\right]\cr
\hspace*{-2.5cm} \mathrm{where}~~~s_2=\left(\frac{\rho_1}{\sqrt{5}}+\frac{\rho_2}{5\sqrt{2}}\right)
\frac{\left[(1-2\rho_0)^2
+3\rho_0(1-\rho_0)+(1-2\rho_0)\sqrt{(1-2\rho_0)^2+6\rho_0(1-\rho_0)}\right]}{\rho_0(1-\rho_0)^2},~~~~~
\label{eq:cur_rev_k_2}
\eea
with $\rho_0=1-\rho_1-\rho_2$. Finally, by replacing $z_{1,2}$ from (\ref{eq:fug_den_k_2}) into (\ref{eq:cur_rev_k_2})-one gets the expression of 
$\langle J\rangle_{I}(\rho_1,\rho_2)$ which is now only a function of the particle densities $\rho_{1,2}$ for the given set of 
values of the rates. 
\begin{figure}
\vspace*{.5 cm}
 \centering
\includegraphics[height=8.5 cm]{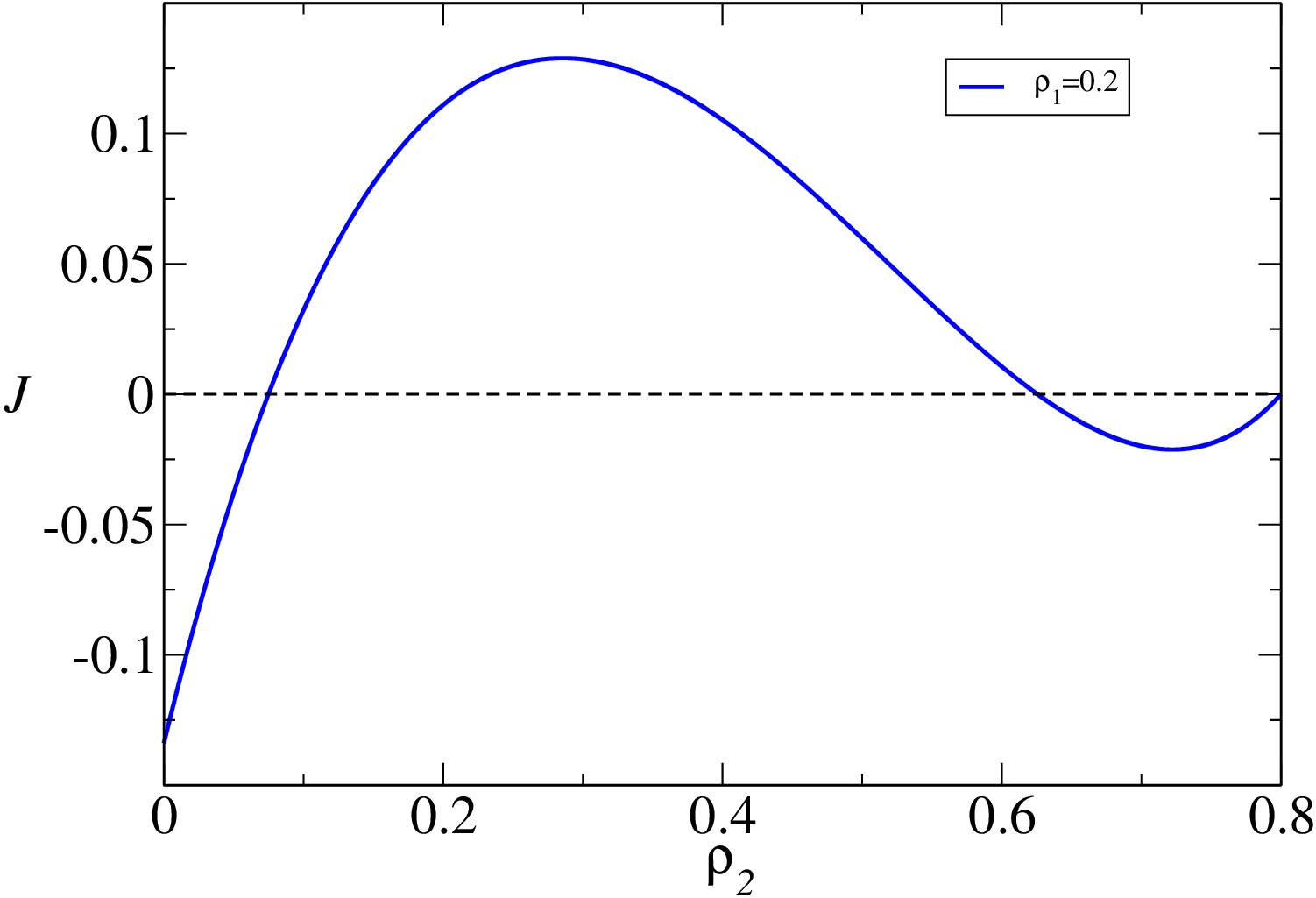}
 \caption{ Density dependent current reversal for a two species assisted hopping and exchange process- multiple(two) point of 
 reversal at $\rho_1=0.2$.}
 \label{fig:3}
\end{figure}
We have plotted this exact expression of the steady state particle current as a function of the second species 
particle density $\rho_2$ for a fixed value $\rho_1=0.2$- the resulting curve is shown in Fig. (\ref{fig:3}). We see that in 
Fig. (\ref{fig:3}), the current reverses its direction twice in contrast to the case $k=1$ in Fig. (\ref{fig:1}) where the 
current changed its direction only once giving rise to a single point of reversal, here in Fig. (\ref{fig:3}), clearly twice change in 
the direction of current results in multiple points of current reversal  which in this case are $\rho_2^\ast\approx0.07$ and 
$\rho_2^\ast\approx0.62$ where the current become zero before reversing its direction. Along with this event of multiple 
reversal points, note that the current in Fig. (\ref{fig:3}) is still nonzero at $\rho_2=0$ because there is a finite number   
of particles of the first species still hopping on the lattice whereas for the $K=1$ case in Fig. (\ref{fig:1}) the current was 
naturally zero at $\rho_1=0$. Also note that this current reversal could also be observed by tuning 
$\rho_1$ with a fixed value of $\rho_2$. In fact, in general, if one plots $\langle J\rangle(\rho_1,\rho_2)$ as a function of both $\rho_1$ 
and $\rho_2$, we may expect a line of reversal in the $\rho_1-\rho_2$ plane along which the current becomes zero before changing its 
direction. 

So, like single species assisted hopping model, we have shown that the emergence of density dependent current reversal can also occur 
for two species assisted hopping and exchange process- this leads us to believe that current reversal can generally happen in multi-
species assisted hopping and exchange models for a broad class of rates.

\subsection{Negative differential mobility in AEM with $K=2$}
\begin{figure}
\vspace*{.5 cm}
 \centering
\includegraphics[height=8.5 cm]{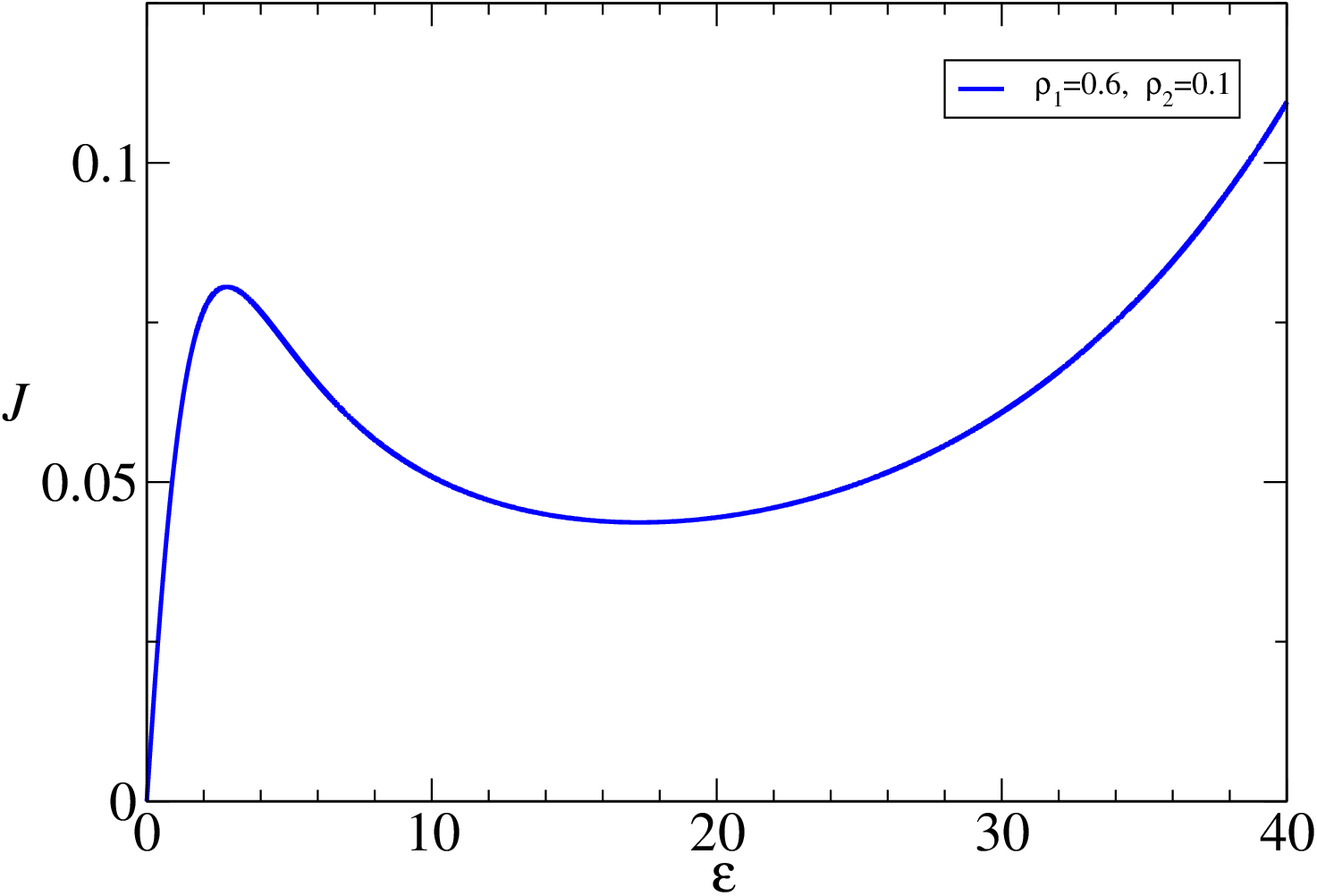}
 \caption{Negative differential mobility exhibited by the total particle current in the region 
 $3 \lessapprox \epsilon \lessapprox 17$ for a two species assisted hopping and exchange model at densities $\rho_1=0.6,\rho_2=0.1$}
 \label{fig:4}
\end{figure}

Just like current reversal, in this section, we would like to show that assisted exchange models 
exhibits  negative differential mobility of particle current.  In particular, we consider $K=2$  
 with  pair-weight functions- 
 \bea
&&g(0,0)=e^{\epsilon},g(1,0)=g(0,1)=e^{\epsilon/2},g(1,1)=(1+\frac{\epsilon}{2}),g(2,0)=e^{\epsilon/5},g(0,2)=e^{7\epsilon/10},\cr&&
g(1,2)=(1+\frac{\epsilon}{2})e^{\epsilon/5},g(2,1)=(1+\frac{\epsilon}{2})e^{-3\epsilon/10},
g(2,2)=(1+\frac{\epsilon}{2})e^{-\epsilon/10}.
\eea
Corresponding  two species assisted exchange dynamics can be  derrived  from
Eq. (\ref{eq:con1}). Since, here we focus on the current as a function of the bias $\epsilon,$ we 
consider the particle densities of the two species to be fixed at $\rho_1=0.6$ and $\rho_2=0.1$ respectively. It is quite straightforward 
to check that all the relevant biases in the total particle current are increasing functions of the bias $\epsilon$ under 
consideration. Now, the exact form of the current in this case can be calculated directly from Eq. (\ref{eq:cur_cond1})- but 
it looks very complicated-so we omit the functional form here. Instead, we show the plot of the total particle current as a function 
of the bias in Fig. (\ref{fig:4}).
In Fig. (\ref{fig:4}), we observe that in the region as we increase the bias, the current decreases indicating the onset of negative 
differential mobility of the particles. Note the difference in the qualitative features of the negative response for the 
single species problem in Fig. (\ref{fig:2})(green curve) with respect to the negative mobility showed by the two species model in 
Fig. (\ref{fig:4})- in the later case, the negative response occurs in a {\it bounded} region value of the bias, more precisely for 
$3 \lessapprox \epsilon \lessapprox 17$ i.e. the current, which at first increases with bias before decreasing in a certain bias value 
region, can increase again with bias beyond that region  whereas in Fig. (\ref{fig:2}) once the current starts decreasing with increasing bias-it 
never goes up again resulting in no upper bound in region of bias causing negative response which is $4\lessapprox \epsilon < \infty$. \\

.

\section{Conclusion}
We have introduced an assisted hopping and exchange model on a one dimensional periodic lattice with hard core particles of $k$
(any finite positive integer) different species along with vacancies- the dynamics conserves the total number of particles as well 
as the number of particles of each species. The model is called {\it assisted} because the hop rate of any kind of particle to 
a vacant nearest neighbor or the exchange rate between two different kind of particles- both explicitly depend on the presence of 
the particle type or vacancy in the neighboring sites other than the sites involved in the hopping or exchange process. For example, 
we considered a single species assisted hopping model(obviously there is no exchange here) where the isolated particles(having 
both neighbors vacant) and crowded particles(having one neighbor vacant) move with different rates. Firstly, we investigate the 
possible steady state measure of this stochastic process and obtained that a pair-factorized steady state is indeed possible for 
two very different broad class of dynamics. Since the pair factorized state generates spatial correlation by definition, we 
opt for a transfer matrix formalism that helps us to calculate the spatial correlation function and expectation values of 
several other observables. We are mostly interested in the particle current that characterizes a non-equilibrium state- so we 
have derived exact expressions for the current in the corresponding PFSS of both the dynamics. To illustrate intriguing features 
like density dependent current reversal for a fixed set of rates and negative differential response of the particle current with 
increasing bias, we have extensively discussed a single species assisted hopping model(where isolated and crowded particles hop 
with different rates) that resembles to partially asymmetric conserved lattice gas \cite{asep_9}. Moreover, these interesting events of 
current reversal and negative differential mobility has also been described briefly for two species assisted hopping and exchange 
model with specified rates, where we observe additional features like {\it multiple} points of reversal in context of current reversal and an  unusual  response  in particle current  where  current as a function of bias  shows 
two extrema.

We conclude that the phenomena like density dependent reversal of current and negative differential mobility of particles can 
generally occur in multi-species assisted hopping and exchange models for suitable choice of the dynamical rates. One can in general ask if AEM  can have a   steady state measure, other  than  PFSS and  explore the possibility of phase separated states.  It would  be interesting to explore these exactly solvable models to study the characterization of non-equilibrium states in terms of current and its higher order cumulants.


\begin{thebibliography}{99}


\bibitem{book1} {\it Nonequilibrium Statistical Mechanics In One Dimension}, ed. V. Privman, 1997 Cambridge
University Press, Cambridge.

\bibitem{cur_press} B. Derrida and J. L. Lebowitz, Phys. Rev. Lett. {\bf 80}, 209 (1998).

\bibitem{cfluc_phse_tran} A. Gerschenfeld and B. Derrida, EPL {\bf 96}, 20001 (2011).

\bibitem{cur_universal} E. Akkermans, T. Bodineau, B. Derrida and O. Shpielberg, EPL {\bf 103}, 20001 (2013).

\bibitem{crev_azrp} A K. Chatterjee and P. K. Mohanty, J. Stat. Mech., {\bf 2017} 093201 (2017).

\bibitem{ndm_1}  R. K. P.~Zia,  E.~L.~Pr{\ae}stgaard, and O.G.~Mouritsen,Am. J. Phys. {\bf 70}, 384 (2002).

\bibitem{ndm_2} A. S. Maksimenko and G. Ya. Slepyan, Phys. Rev. Lett. {\bf 84}, 362 (2000). 

\bibitem{ndm_3} P. Baerts, U. Basu, C. Maes, S. Safaverdi, Phys. Rev. E {\bf 88}, 052109 (2013). 

\bibitem{ndm_4} A. K. Chatterjee, U. Basu and P. K. Mohanty, Phys. Rev. E {\bf 97}, 052137(2018). 

\bibitem{anm_1} R. Eichhorn, P. Reimann, and P. H\"{a}nggi, Phys. Rev. Lett. {\bf 88}, 190601 (2002).
{\it ibid},  ̈Phys. Rev. E {\bf 66}, 066132 (2002).

\bibitem{anm_2} A. Ros, R. Eichhorn, J. Regtmeier, T. T. Duong, P. Reimann and D. Anselmetti,
Nature {\bf 436} 928 (2005).

\bibitem{anm_3} J. Cividini, D. Mukamel, and H. A. Posch, J. Phys. A: Math. Theor. {\bf 51}, 085001 (2018).

\bibitem{book2} J. Marro, R. Dickman, {\it Nonequilibrium Phase Transitions in  Lattice  Models},  1999 Cambridge  University  Press,  New York.

\bibitem{book3} B. Schmittmann and R. K. P. Zia, {\it Statistical Mechanics of Driven Diffusive Systems}, ed. C. Domb 
and J. L. Lebowitz, 1995 Academic Press, New York.

\bibitem{asep_1}   G. M. Sch\"utz,  {\it Phase  Transitions  and  Critical  Phenomena} Vol. 19, ed. C. Domb and J. L. Lebowitz 
2000 Academic Press, London.

\bibitem{asep_2} B. Derrida, E. Domany and D. Mukamel, J. Stat. Phys. {\bf 69}, 667 (1992).

\bibitem{asep_3} S. Prolhac and K. Mallick, J. Phys. A: Math. Theor. {\bf 41(17)}, 175002 (2008).

\bibitem{book4}  T.  M.  Liggett, {\it Stochastic  Interacting  Systems:   Voter, Contact  and  Exclusion  Processes}, 1999 Springer, New York.

\bibitem{asep_4} S. Prolhac and K. Mallick, J. Phys. A: Math. Theor. {\bf 41(36)}, 365003 (2008).

\bibitem{asep_5} M. Gorissen, A. Lazarescu, K. Mallick and C. Vanderzande, Phys. Rev. Lett. {\bf 109}, 170601 (2012).

\bibitem{asep_6} A. Lazarescu,  J. Phys. A: Math. Theor. {\bf 46}, 145003 (2013). 

\bibitem{asep_7} B. Derrida, J. Stat. Mech.,  P07023 (2007).

\bibitem{asep_8} M.R. Evans, Europhys. Lett. {\bf 36}, 13 (1996).

\bibitem{asep_9} U. Basu and P. K. Mohanty, Phys. Rev. E {\bf 79}, 041143 (2009). 



\bibitem{Isaev} A. P. Isaev, P. N. Pyatov and V Rittenberg, Phys. A: Math. Gen. {\bf 34},  5815 (2001)

\bibitem{ZRP_rev} M. R. Evans, T. Hanney, J. Phys. A: Math. Gen. {\bf 38},  R195(2005). 

\bibitem{ABC1}  M. Clincy, B. Derrida, and M. R. Evans, Phys. Rev. E {\bf 67}, 066115 (2003).
 
 \bibitem{ABC2} A. Ayyer, E. A. Carlen, J. L. Lebowitz,P. K. Mohanty, D. Mukamel and E. R. Speer, J. Stat. Phys {\bf 137}, 1166 (2009).

 \bibitem{AHR} P.F. Arndt, T. Heinzel, and V. Rittenberg, J. Phys A: Math. Gen. {\bf 31}, L45 (1998); {\it ibid,} J. Stat. Phys. {\bf 97}, 1 (1999).
 
 \bibitem{MPA}  R. A. Blythe, M. R. Evans, J. Phys. A: Math. Theor. {\bf 40},  R333 (2007). 
 
 \bibitem{EKLS} Y. Kafri, E. Levine, D. Mukamel, G.M. Sch ̈\"utz, and J.
T\"or\"ok, Phys. Rev. Lett. {\bf 89}, 035702 (2002);
M. R. Evans, E. Levine, P. K. Mohanty, D. Mukamel, Euro. Phys. J. B {\bf 41}, 223 (2004).

\bibitem{AF}A. Kundu and  P. K. Mohanty, Physica A {\bf 390}, 1585 (2011).



 
 


\end{thebibliography}
\end{document}